\documentclass[prb,amssymb,twocolumn,superscriptaddress,showpacs]{revtex4}

\usepackage{graphicx}

\usepackage{amsmath}

\usepackage{bm}

\begin{document}

\title{Short time relaxation of a driven elastic string in a random medium}

\author{Alejandro B. Kolton} \affiliation{DPMC-MaNEP, University of
Geneva, 24 Quai Ernest Ansermet, CH-1211 Geneva 4, Switzerland}

\author{Alberto Rosso} \affiliation{LPTMS; CNRS and Universit\'e
Paris-Sud, UMR 8626, ORSAY CEDEX 91405, France}

\author{Ezequiel V. Albano} \affiliation{INIFTA, Facultad C. Exactas UNLP, CONICET, Casilla de Correo 16, Sucursal 4, (1900) La Plata, Argentina}

\author{Thierry Giamarchi} \affiliation{DPMC-MaNEP, University of
Geneva, 24 Quai Ernest Ansermet, CH-1211 Geneva 4, Switzerland}

\begin{abstract}
We study numerically the relaxation of a driven elastic string in a
two dimensional pinning landscape. The relaxation of the string,
initially flat, is governed by a growing length $L(t)$ separating the
short steady-state equilibrated lengthscales, from the large
lengthscales that keep memory of the initial condition. We find a
macroscopic short time regime where relaxation is universal, both
above and below the depinning threshold, different from the one
expected for standard critical phenomena. Below the threshold, the
zero temperature relaxation towards the first pinned configuration
provides a novel, experimentally convenient way to access all the
critical exponents of the depinning transition independently.
\end{abstract}


\maketitle
The study of the dynamics of elastic interfaces in disordered media is
relevant for diverse experimental situations ranging from magnetic
\cite{lemerle_domainwall_creep,shibauchi_creep_magnetic,caysol_minibridge_domainwall}
or ferroelectric \cite{tybell_ferro_creep,paruch_2.5} domain walls,
contact lines of liquid menisci on a rough substrate
\cite{moulinet_rosso}, to crack propagation
\cite{maloy,ponson_fracture}. A fundamental problem is the response of
these systems to external fields which pull the elastic interface with
a force $f$.

A considerable progress has been done in understanding the steady
state dynamics under the applied force. At zero temperature, the
system is pinned by the disorder and the velocity of the interface
remains zero up to a critical force $f_c$.  Above $f_c$ the system
undergoes a depinning transition
\cite{narayan_fisher_cdw,nattermann_stepanow_depinning,chauve_zeta_twoloops,rosso_hartmann,duemmer}
and moves with a non-zero average velocity. Fisher first viewed the
depinning transition as a critical phenomenon
\cite{fisher_depinning_meanfield}: the driving force $f$ plays the
role of the control parameter, the mean velocity $v$ is the order
parameter vanishing at $f_c$ with an exponent $\beta$, and the
divergent correlation length $\xi_f \sim (f-f_c)^{-\nu}$ can be
defined from the velocity-velocity correlation function
\cite{duemmer}. The analogy of the depinning transition with standard
critical phenomena has been however recently challenged. By studying
the low temperature limit of the steady state motion it was found that
no divergent lengthscale exists below threshold
\cite{kolton_vmccreep}.

The non-steady dynamics, although experimentally relevant, has
received less attention. Recently, Schehr and Le Doussal
\cite{schehr_frg} have investigated this regime for an interface
initially flat by analyzing two time correlation functions, as $f
\rightarrow f_c^+$. Their functional renormalization group
calculations show that the transient dynamics displays universal
behavior. Below threshold, numerical studies of the zero temperature
relaxation towards the pinned state have identified a length scale
$\xi_f$ diverging with the exponent $\nu$ at $f_c$
\cite{middleton_narayan,chen_marchetti}. In
Ref.~\onlinecite{kolton_vmccreep} it was shown that this length
$\xi_f$ does not affect steady state properties, but describes
transient processes (deterministic avalanches triggered by thermally
activated events) during the steady state low temperature motion for
$f<f_c$, and it is ultimately related to the vanishing of the
density of metastable states as one approaches $f_c$. The $f=0$
relaxation towards equilibrium at finite temperature has been
studied both numerically
\cite{barrat_vs_yoshino,yoshino_aging_line,kolton_relax,bustingorry_relax,bustingorry_vortexglass}
and analytically \cite{cugliandolo_relaxmanifold,balents_tbl}.

In the present paper we analyze the transient dynamics of an
interface pulled with a finite force close to $f_c$, and show that
it is a powerful method to extract the critical properties of the
depinning transition. Indeed, the analysis of the relaxation
dynamics has been used extensively to study equilibrium critical
phenomena \cite{zheng_std}. The basic idea behind this dynamic
approach is the existence of a growing length $L(t)$.  Let us use
for simplicity the example of the Ising model: the system is
prepared in the ground state, characterized by a global
magnetization $m=1$, and at $t=0$ it is annealed at a temperature
$T$, close to the critical point $T_c$. The global magnetization $m$
relaxes to its equilibrium value following a time evolution
controlled by $L(t)$: for lengths below $L(t)$ the system is
equilibrated, while for lengths larger that $L(t)$ the system keeps
memory of the $t=0$ initial condition. After a microscopic time, the
relaxation is governed by the dynamical exponent $z$ and $L(t) \sim
t^{1/z}$ before reaching the equilibrium correlation length. In this
macroscopic time regime scaling arguments lead to a universal
behavior for the relaxation of the order parameter, although the
system is far from its equilibrium state \cite{zheng_std}.
Analogously, if we assume the presence of such a growing length
$L(t)$ in the transient regime of an initially flat driven elastic
interface, the scaling form for the relaxation of the velocity is
given by
\begin{equation}
v(t,f)= \xi_f^{-\beta/\nu} F(L(t)/\xi_f)
\label{std_v}
\end{equation}
where the function $F(s) \sim s^{-\beta/\nu}$ for small $s$.  When
$s \gtrsim 1$, $F(s) \sim \mbox{const.}$ for $f>f_c$, in order to
get the steady state velocity $v \sim (f-f_c)^\beta$; for $f<f_c$,
where the order parameter is zero at $T=0$, $F(s)$ must be modified
to take into account the exponential decay of the velocity. In
standard phase transitions, a scaling form equivalent to
(\ref{std_v}) describes, in general, the evolution of the order
parameter. In this paper we show that the scaling form (\ref{std_v})
describes the relaxation near depinning, but in a non-standard way.
While in standard critical phenomena $\xi_f$ represents the
correlation length on each side of the
transition, for depinning $\xi_f$ represents the steady state
correlation length only above threshold but a purely transient
correlation length below threshold, absent in the steady state
geometry of the line \cite{kolton_vmccreep}. With this
identification of the relevant lengths the analysis of the transient
gives access to all the critical exponents of the
transition. To make the study more transparent we consider only the
case of strictly zero temperature, in which the interface relaxes
deterministically towards a final pinned configuration without
reaching any steady state below threshold, thus probing transient
deterministic dynamics only.
\begin{figure}
 \centerline{\includegraphics[width=8.5cm]{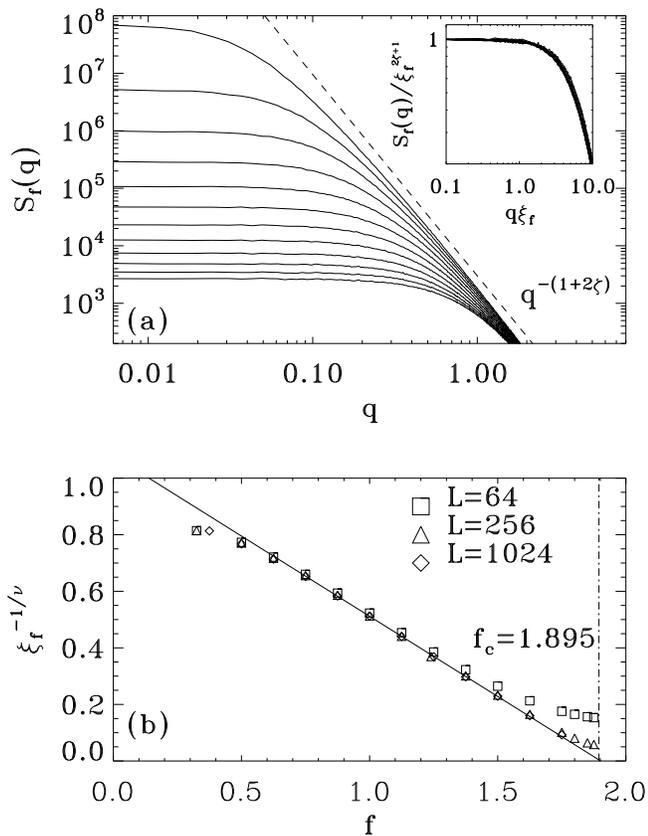}}
 \caption{ (a) Structure factor $S_f(q)$ of the pinned configuration
obtained by relaxing an initially flat line for different forces
$f=0.325,...,2.0$, increasing from the bottom to the top curve.
Inset: collapse using the crossover length $\xi_f$,
and the depinning roughness exponent $\zeta=1.25$. (b) Finite size
study of $\xi_f$ vs $f$. The solid line is a fit to $\xi_f \sim
(f_c-f)^{-\nu}$, where $\nu=1.33$. The dashed-dotted line indicates
the extrapolated value of $f_c$. }
 \label{fig:final_structure}
\end{figure}
We also restrict here to the simple case of a string with short
range elasticity moving in a two dimensional random landscape, but
our method naturally applies to higher dimensional systems as well.

The string is described by a single valued function $u(x,t)$, which
measures its transverse displacement $u$ from the $x$ axis at 
given time $t$. The equation of motion is given by,
\begin{eqnarray}
\partial_t u(x,t) &=&
\partial_x^2 u(x,t) + F_p(u,x) + f
\label{eqmotion}
\end{eqnarray}
where $F_p(u,x)$ is the pinning force with correlations
$\overline{F_p(u,x)F_p(u',x')}=\Delta(u-u')\delta(x-x')$, the
overline represents the average on the disorder realization and
$\Delta(x)$ is a short-ranged function. In our simulations we have
set this range equal to $1$, analyzed system sizes up to $L=2048$,
and results were averaged from $2000$ to $10000$ disorder
realizations. For a fixed force $f<f_c$ we let evolve an initially
flat line $u(x,0)=0$, up to its final pinned configuration, which
can be detected by a very efficient algorithm in polynomial time
\cite{rosso_depinning_simulation}. The geometrical properties of the
line can be described using the averaged structure factor, which for
a general configuration is defined as,
\begin{eqnarray}
S_f(q,t) = \overline{\biggl\langle \biggl|\frac{1}{L}\int_{0}^{L} dz\; u(x,t)
e^{-iqx}\biggr|^2 \biggr\rangle}
\label{S}
\end{eqnarray}
where $L$ is the length of the line, $q=2\pi n/L$, with
$n=1,\ldots,L-1$. For a self affine line we have $S(q) \sim
q^{-(1+2\zeta)}$ thus yielding the roughness exponent $\zeta$.  In
Fig.~\ref{fig:final_structure}(a) we show the structure factor of
the pinned line $S_f(q) \equiv S_f(q,\infty)$ for different values
of $f<f_c$.  Two regimes can be identified: at small lengthscales,
the geometry of the line becomes self-affine and it is characterized
by the depinning roughness exponent $\zeta \approx 1.25$ 
\footnote{Previous studies have found two regimes 
but with an incorrect force dependent
$\zeta$ exponent in the rough regime \cite{vinokur_wrong}.}.  At 
large lengthscales
$S_f(q)$ reaches a plateau which represents the memory of the initial
flat condition. As shown by the perfect collapse in the inset,
$S_f(q)$ is governed by a single length $\xi_f$, given by the
crossover between these two regimes. In
Fig.~\ref{fig:final_structure}(b) we see that the crossover length
increases with the force, diverging with the depinning exponent $\nu$
at a finite value, identified with the threshold $f_c$. Small
deviations from this behavior are observed at very small forces and
whenever the crossover length approaches the system size $L$. Using
the finite-size analysis of Fig.~\ref{fig:final_structure}(b) we can
easily extrapolate the value of the critical force for the infinite
system and identify the crossover length with $\xi_f$. This length
corresponds to the size of the minimal string rearrangement needed to
reach a metastable configuration from the flat one, and its divergence
is due to the vanishing density of metastable states approaching $f_c$
from below. $\xi_f$ can be associated with the divergent length found
in Refs.~\onlinecite{middleton_narayan,chen_marchetti} and with the
size of the deterministic avalanches in the steady state motion
\cite{kolton_vmccreep}. In the following we show that although $\xi_f$
does not affect steady state properties below threshold
\cite{kolton_vmccreep} it affects the transient relaxation.
\begin{figure}
 \centerline{\includegraphics[width=8.5cm]{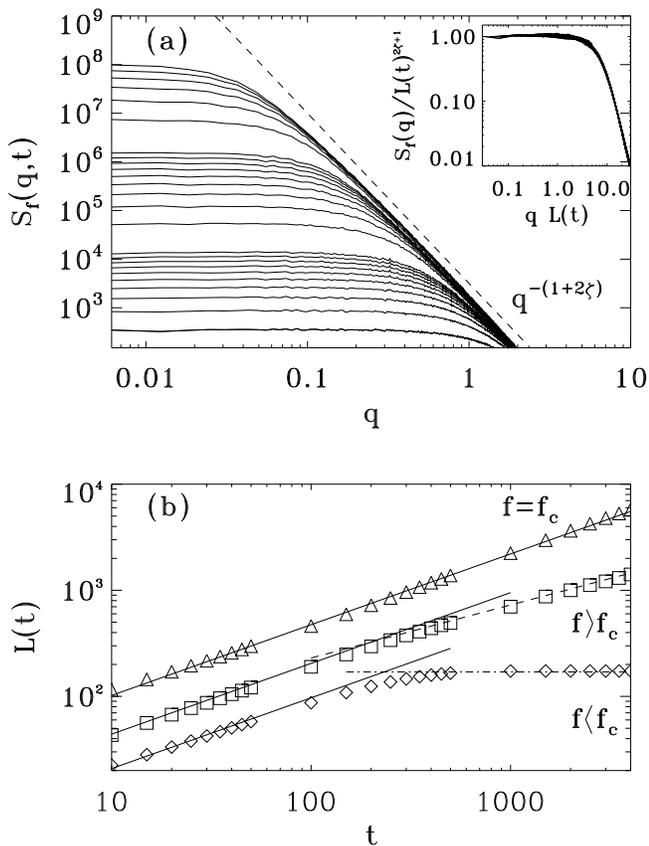}}
 \caption{ (a) Typical evolution of the structure factor $S_f(q,t)$
during the relaxation at $f=1.80$, calculated at different times, increasing
from the bottom to the top curve, obtained by using lines of
size $L=1024$. Inset: collapse using the crossover
length $L(t)$. (b) Growth of $L(t)$ for $f = f_c$ ($\vartriangle$), $f
= 1.75 <f_c$ ($\square$) and $f = 2.05 >f_c$ ($\lozenge$).  The lines
indicate the power law behaviors in the different regimes: $L(t) \sim
t^{2/3}$ (solid), $L(t) \sim t^{1/2}$ (dashed), $L(t)\sim cte.$
(dashed-dotted). The data for each force has been vertically shifted
for clarity.}
 \label{fig:relax_structure}
\end{figure}

Let us now analyze the time evolution of the line towards the final
pinned configuration for $f<f_c$ and to the sliding steady state for
$f>f_c$. For this purpose we solve numerically equation
(\ref{eqmotion}) by using a second-order Runge-Kutta method. In Fig.
\ref{fig:relax_structure} (a) we show the typical evolution of
$S_f(q,t)$ for a force $f=1.80$. Once again, two roughness regimes
are observed, one corresponding to the memory of the flat initial
configuration and the other to the depinning roughness $\zeta=1.25$.
As we can see in the inset, a growing lengthscale, identified with
$L(t)$, governs the time evolution. In
Fig.~\ref{fig:relax_structure} (b) we show the time evolution of
$L(t)$ for forces above and below the threshold. We can distinguish
three regimes for the evolution of $L(t)$. After a first microscopic
time regime where the line is practically flat, we find a
macroscopic short time regime where the growth of $L(t)$ is
controlled by the depinning dynamical exponent $z\sim 1.5$, as
$L(t)\sim t^{1/z}$. This result shows that the depinning transition
is characterized by a universal short time relaxation.  The
crossover to the third regime occurs when $L(t)\sim \xi_f$, after
which we can distinguish the relaxation above or below threshold.
For $f<f_c$, $L(t)$ saturates to $\xi_f$, the characteristic length
of the final pinned configuration. For $f>f_c$, $L(t)$ continues to
grow as $L(t)\sim t^{1/2}$. The thermal dynamical exponent $z=2$ is
produced by the finite velocity, which makes disorder act as an
effective thermal noise above the lengthscale $\xi_f$
\cite{duemmer}. Finite-size effects are expected if one of the two
important length scales, $\xi_f$ and $L(t)$, become of the order of
the system size $L$.
\begin{figure}
 \centerline{\includegraphics[width=8.5cm]{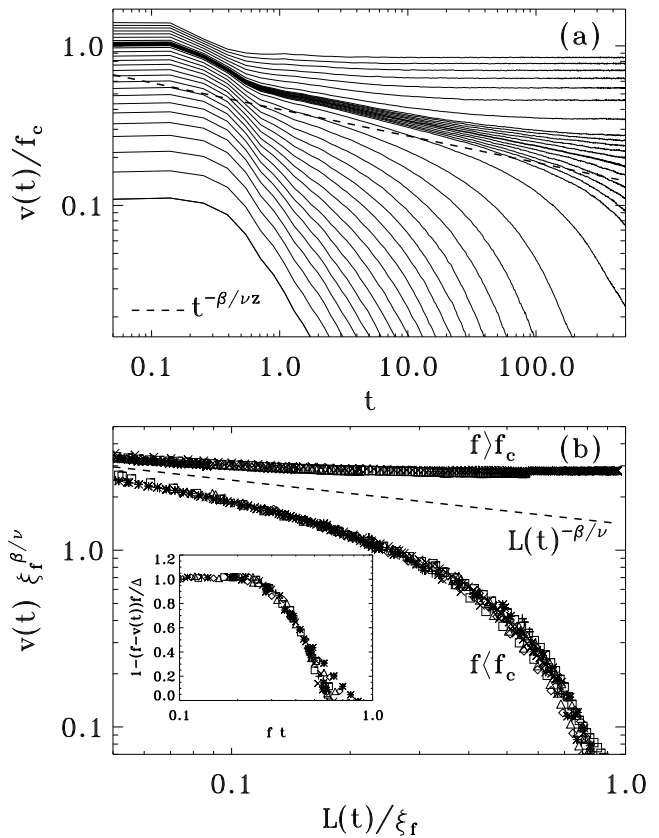}}
 \caption{(a) Relaxation of the velocity $v(t,f)$ for different forces
around the critical force, from $f=0.5$ to $f=2.50$. The dashed
line indicates the power-law behaviour $t^{-\beta/\nu z}$, where
$\beta,\; \nu, \; z$ are critical exponents. (b) Collapse of all
forces using the scaling of Eq.(\ref{std_v}) based on the
existence of a growing length $L(t)$ and a second static
length $\xi_f$ associated with the density of metastable
states. The dashed line indicates the power-law
behaviour $L(t)^{-\beta/\nu}$.
The inset shows the scaling in the
microscopic time regime, as predicted by
the perturbation theory (Eq(\ref{micro})).}
 \label{fig:relax_velocity}
\end{figure}

In the steady state the observables are controlled by the
correlation length $\xi_f$. In particular, the mean velocity, order
parameter of the depinning transition, is given by $v \sim
\xi_f^{-\beta/\nu}$. However, in the transient regime, $L(t)$ plays
a crucial role in the relaxation of all the observables. In
particular, in Fig.\ref{fig:relax_velocity} (a) we see that the
three growth regimes previously found for $L(t)$ correspond to three
regimes of $v(t,f)$. At the shortest times the line moves freely, $v
\sim f$, since pinning forces average to zero for a flat line. The
first non zero corrections to this free-flow behavior can be
therefore obtained by using a second order perturbation theory in
the disorder. This gives,
\begin{equation}
v(t,f) \sim f - \frac{\Delta(0)-\Delta(ft)}{f}.
\label{micro}
\end{equation}
The inset of Fig.~\ref{fig:relax_velocity}(b) shows that (\ref{micro})
holds up to times $t < r_f/f$, where $r_f=1$ is the correlation length
of the disorder. The first stage of microscopic relaxation of $v(t,f)$
gives thus access to the shape of the disorder correlator $\Delta(u)$
for $u < r_f$. The data obtained at longer times verify the scaling
relation proposed in (\ref{std_v}), as we show in
Fig.~\ref{fig:relax_velocity}(b). This collapse shows that the
velocity is controlled by the ratio $L(t)/\xi_f$.  When $L(t) \ll
\xi_f$ the growing length $L(t)$ is the only relevant scale in the
problem and replaces $\xi_f$ in the steady state relation $v \sim
\xi_f^{-\beta/\nu}$. This leads to $v(t,f) \sim L(t)^{-\beta/\nu}$, as
it is observed in the initial slope of the collapsed curves in
Fig.~\ref{fig:relax_velocity}(b), with $\beta \sim
0.33$. Subsequently, for $f>f_c$ the scaling function tends to a
constant while for $f<f_c$ has a fast decay.  A similar relaxation of
the velocity was found in other disordered models 
\cite{lee_shortime,zapperi}.

In conclusion, we have identified the relevant static and dynamical
lengths, $L(t)$ and $\xi_f$, controlling the dynamics of relaxation
of an elastic line under an applied force. We have shown that
(\ref{std_v}) describes well the universal relaxation of the
velocity observed for $t>1/f$ and $f$ close to the threshold. In
contrast to standard critical phenomena, below threshold the static
length $\xi_f$ does not represent any steady state correlation
length. The proposed scaling relationship for the time relaxation
can be used to get all the critical exponents of the depinning
transition independently: we can get $\nu$ from the characteristic
length $\xi_f$ of the pinned configuration below threshold (or from
the crossover to the thermal growth of $L(t)$ above threshold), $z$
from the evolution of $L(t)$, and $\beta$ from the relaxation of the
velocity. This is best accomplished close to the depinning threshold
since $\xi_f$ is large, and the macroscopic short time regime
longer. As for equilibrium critical phenomena, this is a convenient
method for both, numerics and experiments, since it avoids the
problem of equilibrating a sample in the steady state near the
depinning threshold. When $L(t)$ becomes of the order of the
correlation length $\xi_f$ all the observables approach their steady
state value for $f>f_c$ or their value at the final pinned
configuration for $f<f_c$ exponentially fast. This study is
therefore relevant for experimental situations whenever $T$ is low
enough to assure well separated time scales for deterministic and
thermally activated motion.

Finally, we point out that the experimental characterization of
interfaces pinned below $f_c$, prepared under a unique protocol,
should be a convenient tool to investigate depinning, since the
geometry of these interfaces yields an independent measure of the
exponents $\zeta$ and $\nu$, as shown in
Fig.~\ref{fig:final_structure}.  This feature is important, as these
exponents are usually related.  The so called statistical tilt
symmetry (STS) is present whenever the elastic force is a linear
function of the deformation field $u(x)$. In general, the STS
relation depends on the range of the elastic interactions: for a
short-ranged elasticity, as in (\ref{eqmotion}), the relation reads
$\nu=1/(2-\zeta)$, while for the long-ranged interactions, expected
for crack propagation in solids or for the contact line of liquids,
the relation becomes $\nu=1/(1-\zeta)$. Recent experiments in these
two systems \cite{maloy,moulinet_contact_line} gave roughness
exponents $\zeta$ that are systematically bigger than the one
computed by numerical simulations \cite{rosso_depinning_simulation}.
In order to explain these deviations the presence of non-linear
elastic corrections able to change the universal behavior of the
depinning transition
\cite{rosso_depinning_simulation,ledoussal_kpzinfracture,eytan_longrange}
has been invoqued. The measurement of an eventual violation of the
statistical tilt symmetry relation from the study of interfaces
pinned below $f_c$ could be thus used as a novel smoking gun test
for this hypothesis.

We acknowledge discussions with O. D\"{u}mmer, W. Krauth and P. Le
Doussal. This work was supported in part by the Swiss National Fund
under MaNEP and Division II and by NSF grant PHY99-07949.


\begin{thebibliography}{33}
\expandafter\ifx\csname natexlab\endcsname\relax\def\natexlab#1{#1}\fi
\expandafter\ifx\csname bibnamefont\endcsname\relax
  \def\bibnamefont#1{#1}\fi
\expandafter\ifx\csname bibfnamefont\endcsname\relax
  \def\bibfnamefont#1{#1}\fi
\expandafter\ifx\csname citenamefont\endcsname\relax
  \def\citenamefont#1{#1}\fi
\expandafter\ifx\csname url\endcsname\relax
  \def\url#1{\texttt{#1}}\fi
\expandafter\ifx\csname urlprefix\endcsname\relax\def\urlprefix{URL }\fi
\providecommand{\bibinfo}[2]{#2}
\providecommand{\eprint}[2][]{\url{#2}}

\bibitem[{\citenamefont{Lemerle et~al.}(1998)\citenamefont{Lemerle, Ferr{\'e},
  Chappert, Mathet, Giamarchi, and {Le Doussal}}}]{lemerle_domainwall_creep}
\bibinfo{author}{\bibfnamefont{S.}~\bibnamefont{Lemerle}},
  \bibinfo{author}{\bibfnamefont{J.}~\bibnamefont{Ferr{\'e}}},
  \bibinfo{author}{\bibfnamefont{C.}~\bibnamefont{Chappert}},
  \bibinfo{author}{\bibfnamefont{V.}~\bibnamefont{Mathet}},
  \bibinfo{author}{\bibfnamefont{T.}~\bibnamefont{Giamarchi}},
  \bibnamefont{and} \bibinfo{author}{\bibfnamefont{P.}~\bibnamefont{{Le
  Doussal}}}, \bibinfo{journal}{Phys. Rev. Lett.}
  \textbf{\bibinfo{volume}{80}}, \bibinfo{pages}{849} (\bibinfo{year}{1998}).

\bibitem[{\citenamefont{{Shibauchi {\it et
  al.}}}(2001)}]{shibauchi_creep_magnetic}
\bibinfo{author}{\bibfnamefont{T.}~\bibnamefont{{Shibauchi {\it et al.}}}},
  \bibinfo{journal}{Phys. Rev. Lett.} \textbf{\bibinfo{volume}{87}},
  \bibinfo{pages}{267201} (\bibinfo{year}{2001}).

\bibitem[{\citenamefont{{Caysoll {\it et
  al.}}}(2004)}]{caysol_minibridge_domainwall}
\bibinfo{author}{\bibfnamefont{F.}~\bibnamefont{{Caysoll {\it et al.}}}},
  \bibinfo{journal}{Phys. Rev. Lett.} \textbf{\bibinfo{volume}{92}},
  \bibinfo{pages}{107202} (\bibinfo{year}{2004}).

\bibitem[{\citenamefont{Tybell et~al.}(2002)\citenamefont{Tybell, Paruch,
  Giamarchi, and Triscone}}]{tybell_ferro_creep}
\bibinfo{author}{\bibfnamefont{T.}~\bibnamefont{Tybell}},
  \bibinfo{author}{\bibfnamefont{P.}~\bibnamefont{Paruch}},
  \bibinfo{author}{\bibfnamefont{T.}~\bibnamefont{Giamarchi}},
  \bibnamefont{and} \bibinfo{author}{\bibfnamefont{J.~M.}
  \bibnamefont{Triscone}}, \bibinfo{journal}{Phys. Rev. Lett.}
  \textbf{\bibinfo{volume}{89}}, \bibinfo{pages}{097601}
  (\bibinfo{year}{2002}).

\bibitem[{\citenamefont{Paruch et~al.}(2005)\citenamefont{Paruch, Triscone, and
  Giamarchi}}]{paruch_2.5}
\bibinfo{author}{\bibfnamefont{P.}~\bibnamefont{Paruch}},
  \bibinfo{author}{\bibfnamefont{J.~M.} \bibnamefont{Triscone}},
  \bibnamefont{and}
  \bibinfo{author}{\bibfnamefont{T.}~\bibnamefont{Giamarchi}},
  \bibinfo{journal}{Phys. Rev. Lett.} \textbf{\bibinfo{volume}{94}},
  \bibinfo{pages}{197601} (\bibinfo{year}{2005}).

\bibitem[{\citenamefont{Moulinet et~al.}(2004)\citenamefont{Moulinet, Rosso,
  Krauth, and Rolley}}]{moulinet_rosso}
\bibinfo{author}{\bibfnamefont{S.}~\bibnamefont{Moulinet}},
  \bibinfo{author}{\bibfnamefont{A.}~\bibnamefont{Rosso}},
  \bibinfo{author}{\bibfnamefont{W.}~\bibnamefont{Krauth}}, \bibnamefont{and}
  \bibinfo{author}{\bibfnamefont{E.}~\bibnamefont{Rolley}},
  \bibinfo{journal}{Phys. Rev. E} \textbf{\bibinfo{volume}{69}},
  \bibinfo{pages}{035103} (\bibinfo{year}{2004}).

\bibitem[{\citenamefont{M{\aa}l{\o}y and Schmittbuhl}(2001)}]{maloy}
\bibinfo{author}{\bibfnamefont{K.~J.} \bibnamefont{M{\aa}l{\o}y}}
  \bibnamefont{and}
  \bibinfo{author}{\bibfnamefont{J.}~\bibnamefont{Schmittbuhl}},
  \bibinfo{journal}{Phys. Rev. Lett.} \textbf{\bibinfo{volume}{87}},
  \bibinfo{pages}{105502} (\bibinfo{year}{2001}).

\bibitem[{\citenamefont{Ponson et~al.}(2006)\citenamefont{Ponson, Bonamy, and
  Bouchaud}}]{ponson_fracture}
\bibinfo{author}{\bibfnamefont{L.}~\bibnamefont{Ponson}},
  \bibinfo{author}{\bibfnamefont{D.}~\bibnamefont{Bonamy}}, \bibnamefont{and}
  \bibinfo{author}{\bibfnamefont{E.}~\bibnamefont{Bouchaud}},
  \bibinfo{journal}{Phys. Rev. Lett.} \textbf{\bibinfo{volume}{96}},
  \bibinfo{pages}{035506} (\bibinfo{year}{2006}).

\bibitem[{\citenamefont{Narayan and Fisher}(1992)}]{narayan_fisher_cdw}
\bibinfo{author}{\bibfnamefont{O.}~\bibnamefont{Narayan}} \bibnamefont{and}
  \bibinfo{author}{\bibfnamefont{D.~S.} \bibnamefont{Fisher}},
  \bibinfo{journal}{Phys. Rev. B} \textbf{\bibinfo{volume}{46}},
  \bibinfo{pages}{11520} (\bibinfo{year}{1992}).

\bibitem[{\citenamefont{Nattermann et~al.}(1992)\citenamefont{Nattermann,
  Stepanow, Tang, and Leschhorn}}]{nattermann_stepanow_depinning}
\bibinfo{author}{\bibfnamefont{T.}~\bibnamefont{Nattermann}},
  \bibinfo{author}{\bibfnamefont{S.}~\bibnamefont{Stepanow}},
  \bibinfo{author}{\bibfnamefont{L.~H.} \bibnamefont{Tang}}, \bibnamefont{and}
  \bibinfo{author}{\bibfnamefont{H.}~\bibnamefont{Leschhorn}},
  \bibinfo{journal}{J. Phys. (Paris)} \textbf{\bibinfo{volume}{2}},
  \bibinfo{pages}{1483} (\bibinfo{year}{1992}).

\bibitem[{\citenamefont{Chauve et~al.}(2001)\citenamefont{Chauve, Wiese, and
  {Le Doussal}}}]{chauve_zeta_twoloops}
\bibinfo{author}{\bibfnamefont{P.}~\bibnamefont{Chauve}},
  \bibinfo{author}{\bibfnamefont{K.}~\bibnamefont{Wiese}}, \bibnamefont{and}
  \bibinfo{author}{\bibfnamefont{P.}~\bibnamefont{{Le Doussal}}},
  \bibinfo{journal}{Phys. Rev. Lett.} \textbf{\bibinfo{volume}{86}},
  \bibinfo{pages}{1785} (\bibinfo{year}{2001}).

\bibitem[{\citenamefont{Rosso et~al.}(2003)\citenamefont{Rosso, Hartmann, and
  Krauth}}]{rosso_hartmann}
\bibinfo{author}{\bibfnamefont{A.}~\bibnamefont{Rosso}},
  \bibinfo{author}{\bibfnamefont{A.~K.} \bibnamefont{Hartmann}},
  \bibnamefont{and} \bibinfo{author}{\bibfnamefont{W.}~\bibnamefont{Krauth}},
  \bibinfo{journal}{Phys. Rev. E} \textbf{\bibinfo{volume}{67}},
  \bibinfo{pages}{021602} (\bibinfo{year}{2003}).

\bibitem[{\citenamefont{D\"{u}emmer and Krauth}(2005)}]{duemmer}
\bibinfo{author}{\bibfnamefont{O.}~\bibnamefont{D\"{u}emmer}} \bibnamefont{and}
  \bibinfo{author}{\bibfnamefont{W.}~\bibnamefont{Krauth}},
  \bibinfo{journal}{Phys. Rev. E} \textbf{\bibinfo{volume}{71}},
  \bibinfo{pages}{061601} (\bibinfo{year}{2005}).

\bibitem[{\citenamefont{Fisher}(1985)}]{fisher_depinning_meanfield}
\bibinfo{author}{\bibfnamefont{D.~S.} \bibnamefont{Fisher}},
  \bibinfo{journal}{Phys. Rev. B} \textbf{\bibinfo{volume}{31}},
  \bibinfo{pages}{1396} (\bibinfo{year}{1985}).

\bibitem[{\citenamefont{Kolton et~al.}(2006)\citenamefont{Kolton, Rosso,
  Giamarchi, and Krauth}}]{kolton_vmccreep}
\bibinfo{author}{\bibfnamefont{A.~B.} \bibnamefont{Kolton}},
  \bibinfo{author}{\bibfnamefont{A.}~\bibnamefont{Rosso}},
  \bibinfo{author}{\bibfnamefont{T.}~\bibnamefont{Giamarchi}},
  \bibnamefont{and} \bibinfo{author}{\bibfnamefont{W.}~\bibnamefont{Krauth}},
  \bibinfo{journal}{Phys. Rev. Lett.} \textbf{\bibinfo{volume}{97}},
  \bibinfo{pages}{057001} (\bibinfo{year}{2006}).

\bibitem[{\citenamefont{Schehr and {Le Doussal}}(2005)}]{schehr_frg}
\bibinfo{author}{\bibfnamefont{G.}~\bibnamefont{Schehr}} \bibnamefont{and}
  \bibinfo{author}{\bibfnamefont{P.}~\bibnamefont{{Le Doussal}}},
  \bibinfo{journal}{Europhys. Lett.} \textbf{\bibinfo{volume}{71}},
  \bibinfo{pages}{290} (\bibinfo{year}{2005}).

\bibitem[{\citenamefont{Narayan and Middleton}(1994)}]{middleton_narayan}
\bibinfo{author}{\bibfnamefont{O.}~\bibnamefont{Narayan}} \bibnamefont{and}
  \bibinfo{author}{\bibfnamefont{A.~A.} \bibnamefont{Middleton}},
  \bibinfo{journal}{Phys. Rev. B} \textbf{\bibinfo{volume}{49}},
  \bibinfo{pages}{244} (\bibinfo{year}{1994}).

\bibitem[{\citenamefont{Chen and Marchetti}(1995)}]{chen_marchetti}
\bibinfo{author}{\bibfnamefont{L.~W.} \bibnamefont{Chen}} \bibnamefont{and}
  \bibinfo{author}{\bibfnamefont{M.~C.} \bibnamefont{Marchetti}},
  \bibinfo{journal}{Phys. Rev. B} \textbf{\bibinfo{volume}{51}},
  \bibinfo{pages}{6296} (\bibinfo{year}{1995}).

\bibitem[{\citenamefont{Barrat}(1997)}]{barrat_vs_yoshino}
\bibinfo{author}{\bibfnamefont{A.}~\bibnamefont{Barrat}},
  \bibinfo{journal}{Phys. Rev. E} \textbf{\bibinfo{volume}{55}},
  \bibinfo{pages}{5651} (\bibinfo{year}{1997}).

\bibitem[{\citenamefont{Yoshino}(1998)}]{yoshino_aging_line}
\bibinfo{author}{\bibfnamefont{H.}~\bibnamefont{Yoshino}},
  \bibinfo{journal}{Phys. Rev. Lett.} \textbf{\bibinfo{volume}{81}},
  \bibinfo{pages}{1493} (\bibinfo{year}{1998}).

\bibitem[{\citenamefont{Kolton et~al.}(2005)\citenamefont{Kolton, Rosso, and
  Giamarchi}}]{kolton_relax}
\bibinfo{author}{\bibfnamefont{A.~B.} \bibnamefont{Kolton}},
  \bibinfo{author}{\bibfnamefont{A.}~\bibnamefont{Rosso}}, \bibnamefont{and}
  \bibinfo{author}{\bibfnamefont{T.}~\bibnamefont{Giamarchi}},
  \bibinfo{journal}{Phys. Rev. Lett.} \textbf{\bibinfo{volume}{95}},
  \bibinfo{pages}{180604} (\bibinfo{year}{2005}).

\bibitem[{\citenamefont{Bustingorry et~al.}()\citenamefont{Bustingorry, Iguain,
  Chamon, Cugliandolo, and Dominguez}}]{bustingorry_relax}
\bibinfo{author}{\bibfnamefont{S.}~\bibnamefont{Bustingorry}},
  \bibinfo{author}{\bibfnamefont{J.~L.} \bibnamefont{Iguain}},
  \bibinfo{author}{\bibfnamefont{C.}~\bibnamefont{Chamon}},
  \bibinfo{author}{\bibfnamefont{L.~F.} \bibnamefont{Cugliandolo}},
  \bibnamefont{and}
  \bibinfo{author}{\bibfnamefont{D.}~\bibnamefont{Dominguez}},
  \bibinfo{note}{cond-mat/0603503}.

\bibitem[{\citenamefont{Bustingorry et~al.}(2006)\citenamefont{Bustingorry,
  Cugliandolo, and Dominguez}}]{bustingorry_vortexglass}
\bibinfo{author}{\bibfnamefont{S.}~\bibnamefont{Bustingorry}},
  \bibinfo{author}{\bibfnamefont{L.~F.} \bibnamefont{Cugliandolo}},
  \bibnamefont{and}
  \bibinfo{author}{\bibfnamefont{D.}~\bibnamefont{Dominguez}},
  \bibinfo{journal}{Phys. Rev. Lett.} \textbf{\bibinfo{volume}{96}},
  \bibinfo{pages}{027001} (\bibinfo{year}{2006}).

\bibitem[{\citenamefont{Cugliandolo et~al.}(1996)\citenamefont{Cugliandolo,
  Kurchan, and {Le Doussal}}}]{cugliandolo_relaxmanifold}
\bibinfo{author}{\bibfnamefont{L.}~\bibnamefont{Cugliandolo}},
  \bibinfo{author}{\bibfnamefont{J.}~\bibnamefont{Kurchan}}, \bibnamefont{and}
  \bibinfo{author}{\bibfnamefont{P.}~\bibnamefont{{Le Doussal}}},
  \bibinfo{journal}{Phys. Rev. Lett.} \textbf{\bibinfo{volume}{76}},
  \bibinfo{pages}{2390} (\bibinfo{year}{1996}).

\bibitem[{\citenamefont{Balents and {Le Doussal}}(2004)}]{balents_tbl}
\bibinfo{author}{\bibfnamefont{L.}~\bibnamefont{Balents}} \bibnamefont{and}
  \bibinfo{author}{\bibfnamefont{P.}~\bibnamefont{{Le Doussal}}},
  \bibinfo{journal}{Phys. Rev. E} \textbf{\bibinfo{volume}{69}},
  \bibinfo{pages}{061107} (\bibinfo{year}{2004}).

\bibitem[{\citenamefont{Zheng}(2005)}]{zheng_std}
\bibinfo{author}{\bibfnamefont{B.}~\bibnamefont{Zheng}}, in
  \emph{\bibinfo{booktitle}{Computer Simulation Studies in Condensed-Matter
  Physics XVII}}, edited by \bibinfo{editor}{\bibfnamefont{D.~P.}
  \bibnamefont{Landau}}, \bibinfo{editor}{\bibfnamefont{S.~P.}
  \bibnamefont{Lewis}}, \bibnamefont{and} \bibinfo{editor}{\bibfnamefont{H.-B.}
  \bibnamefont{Sch\"{u}ttler}} (\bibinfo{publisher}{Springer-Verlag},
  \bibinfo{address}{Heilderberg}, \bibinfo{year}{2005}).

\bibitem[{\citenamefont{Rosso and Krauth}(2002)}]{rosso_depinning_simulation}
\bibinfo{author}{\bibfnamefont{A.}~\bibnamefont{Rosso}} \bibnamefont{and}
  \bibinfo{author}{\bibfnamefont{W.}~\bibnamefont{Krauth}},
  \bibinfo{journal}{Phys. Rev. E} \textbf{\bibinfo{volume}{65}},
  \bibinfo{pages}{025101R} (\bibinfo{year}{2002}).

\bibitem[{\citenamefont{Lee and Kim}(2006)}]{lee_shortime}
\bibinfo{author}{\bibfnamefont{C.}~\bibnamefont{Lee}} \bibnamefont{and}
  \bibinfo{author}{\bibfnamefont{J.~M.} \bibnamefont{Kim}},
  \bibinfo{journal}{Phys. Rev. E} \textbf{\bibinfo{volume}{73}},
  \bibinfo{pages}{016140} (\bibinfo{year}{2006}).

\bibitem[{\citenamefont{Zapperi et~al.}(2000)\citenamefont{Zapperi,
  J.~S.~Andrade, and Filho}}]{zapperi}
\bibinfo{author}{\bibfnamefont{S.}~\bibnamefont{Zapperi}},
  \bibinfo{author}{\bibfnamefont{J.}~\bibnamefont{J.~S.~Andrade}},
  \bibnamefont{and} \bibinfo{author}{\bibfnamefont{J.~M.} \bibnamefont{Filho}},
  \bibinfo{journal}{Phys. Rev. B} \textbf{\bibinfo{volume}{61}},
  \bibinfo{pages}{14791} (\bibinfo{year}{2000}).

\bibitem[{\citenamefont{Moulinet et~al.}(2002)\citenamefont{Moulinet, Guthmann,
  and Rolley}}]{moulinet_contact_line}
\bibinfo{author}{\bibfnamefont{S.}~\bibnamefont{Moulinet}},
  \bibinfo{author}{\bibfnamefont{C.}~\bibnamefont{Guthmann}}, \bibnamefont{and}
  \bibinfo{author}{\bibfnamefont{E.}~\bibnamefont{Rolley}},
  \bibinfo{journal}{Eur. Phys. J. E} \textbf{\bibinfo{volume}{8}},
  \bibinfo{pages}{437} (\bibinfo{year}{2002}).

\bibitem[{\citenamefont{{Le Doussal} et~al.}(2006)\citenamefont{{Le Doussal},
  Wiese, Raphael, and Golestanian}}]{ledoussal_kpzinfracture}
\bibinfo{author}{\bibfnamefont{P.}~\bibnamefont{{Le Doussal}}},
  \bibinfo{author}{\bibfnamefont{K.~J.} \bibnamefont{Wiese}},
  \bibinfo{author}{\bibfnamefont{E.}~\bibnamefont{Raphael}}, \bibnamefont{and}
  \bibinfo{author}{\bibfnamefont{R.}~\bibnamefont{Golestanian}},
  \bibinfo{journal}{Phys. Rev. Lett.} \textbf{\bibinfo{volume}{96}},
  \bibinfo{pages}{015702} (\bibinfo{year}{2006}).

\bibitem[{\citenamefont{Adda-Bedia et~al.}(2006)\citenamefont{Adda-Bedia,
  Katzav, and Vandembroucq}}]{eytan_longrange}
\bibinfo{author}{\bibfnamefont{M.}~\bibnamefont{Adda-Bedia}},
  \bibinfo{author}{\bibfnamefont{E.}~\bibnamefont{Katzav}}, \bibnamefont{and}
  \bibinfo{author}{\bibfnamefont{D.}~\bibnamefont{Vandembroucq}},
  \bibinfo{journal}{Phys. Rev. E} \textbf{\bibinfo{volume}{73}},
  \bibinfo{pages}{035106} (\bibinfo{year}{2006}).

\bibitem[{\citenamefont{Leaf. et~al.}(2002)\citenamefont{Leaf., Obukhov,
  Scheidl, and Vinokur}}]{vinokur_wrong}
\bibinfo{author}{\bibfnamefont{G.}~\bibnamefont{Leaf.}},
  \bibinfo{author}{\bibfnamefont{S.}~\bibnamefont{Obukhov}},
  \bibinfo{author}{\bibfnamefont{S.}~\bibnamefont{Scheidl}}, \bibnamefont{and}
  \bibinfo{author}{\bibfnamefont{V.}~\bibnamefont{Vinokur}},
  \bibinfo{journal}{Journal of Magnetism and Magnetic Materials}
  \textbf{\bibinfo{volume}{241}}, \bibinfo{pages}{118} (\bibinfo{year}{2002}).

\end{thebibliography}

\end{document}